\documentclass{aa}  

\usepackage{graphicx}
\usepackage{txfonts}
\usepackage{xcolor}
\usepackage{gensymb}

\usepackage[hyperindex=true, breaklinks=true,colorlinks=true,urlcolor=blue, linkcolor=blue,  citecolor=blue, bookmarks=true, bookmarksopen=true]{hyperref}

\usepackage{enumitem}

\newcommand{\mum}{$\mu$m}

\begin{document}

   \title{A triple star in disarray}

   \subtitle{Multi-epoch observations of T Tauri with VLT-SPHERE and LBT-LUCI}

   \author{M. Kasper
          \inst{1,2}
          \and
          K. K. R. Santhakumari \inst{3,4}
          \and 
          T.M. Herbst \inst{4}
          \and
          R. van Boekel \inst{4}
          \and
          F. Menard \inst{2}
          \and
          R. Gratton \inst{3}
          \and 
          R.G. van Holstein \inst{6,13}
          \and
           M. Langlois \inst{5}
          \and
          C. Ginski \inst{6}
          \and
          A. Boccaletti\inst{12}
          \and
          M. Benisty \inst{2}
          \and
          J. de Boer \inst{6}
          \and
          P. Delorme \inst{2}
          \and
          S. Desidera \inst{3}
          \and
          C. Dominik \inst{7}
          \and
          J. Hagelberg \inst{14}
          \and
          T. Henning \inst{4}
          \and
          Jochen Heidt \inst{16}
          \and
          R. K\"ohler\inst{15}
          \and
          D. Mesa\inst{3}
          \and
          S. Messina\inst{3}
          \and 
          A. Pavlov\inst{4}
          \and
          C. Petit\inst{8}
          \and
          E. Rickman\inst{14}
          \and
          A. Roux\inst{2}
          \and
          F. Rigal\inst{7}
          \and
          A. Vigan\inst{9}
          \and
          Z. Wahhaj\inst{13}
          \and
          A. Zurlo\inst{10,11}
          }

   \institute{European Southern Observatory, Karl-Schwarzschild-Str. 2, 85748 Garching, Germany
              \email{mkasper@eso.org}
         \and
             Univ. Grenoble Alpes, CNRS, IPAG, F-38000 Grenoble, France
         \and
             INAF-Osservatorio Astronomico di Padova, Vicolo dell'Osservatorio 5, 35122 Padova, Italy
         \and
             Max-Planck-Institute for Astronomy (MPIA), K\"onigstuhl~17, 69117 Heidelberg, Germany
         \and
            CRAL, CNRS, Universit\'{e} Lyon 1, Universit\'{e} de Lyon, ENS, 9 avenue Charles Andre, 69561 Saint Genis Laval, France
         \and
            Leiden Observatory, Leiden University, PO Box 9513, 2300 RA Leiden, The Netherlands
         \and
            Anton Pannekoek Institute for Astronomy, Science Park 9, NL-1098XH Amsterdam, The Netherland
         \and
            DOTA, ONERA, Universit\'{e} Paris Saclay, F-91123, Palaiseau France
         \and
            Aix Marseille Universit\'e, CNRS, CNES,  LAM, Marseille, France
         \and
            N\'ucleo de Astronom\'ia, Facultad de Ingenier\'ia y Ciencias, Universidad Diego Portales, Av. Ejercito 441, Santiago, Chile
         \and
            Escuela de Ingenier\'ia Industrial, Facultad de Ingenier\'ia y Ciencias, Universidad Diego Portales, Av. Ejercito 441, Santiago, Chile
         \and
            LESIA, Observatoire de Paris, Universit\'e PSL, CNRS, Sorbonne Universit\'e, Univ. Paris Diderot, Sorbonne Paris Cit\'e, 5 place Jules Janssen, 92195 Meudon, France
         \and
            European Southern Observatory, Alonso de Cordova 3107, Vitacura, Casilla 19001, Santiago, Chile
         \and
            Geneva Observatory, University of Geneva, Chemin des Mailettes 51, 1290 Versoix, Switzerland
         \and
            University of Vienna, Department of Astrophysics, T\"urkenschanzstr.~17 (Sternwarte), 1180 Vienna, Austria
         \and
             Landessternwarte, Zentrum f\"ur Astronomie der Universit\"at Heidelberg, K\"onigstuhl~12, 69117 Heidelberg, Germany  
             }

   \date{Received August 17, 2020; accepted October 22, 2020}

 
  \abstract
   {}
   {T Tauri remains an enigmatic triple star for which neither the evolutionary state of the stars themselves, nor the geometry of the complex outflow system is completely understood. Eight-meter class telescopes equipped with state-of-the-art adaptive optics provide the spatial resolution necessary to trace tangential motion of features over a timescale of a few years, and they help to associate them with the different outflows.}
   {We used \textit{J}-, \textit{H}-, and \textit{K}-band high-contrast coronagraphic imaging with VLT-SPHERE recorded between 2016 and 2018 to map reflection nebulosities and obtain high precision near-infrared (NIR) photometry of the triple star. We also present H$_2$ emission maps of the $\nu$ = 1-0 S(1) line at 2.122~\mum{} obtained with LBT-LUCI during its commissioning period at the end of 2016.}
   {The data reveal a number of new features in the system, some of which are seen in reflected light and some are seen in H$_2$ emission; furthermore, they can all be associated with the main outflows. The tangential motion of the features provides compelling evidence that T Tauri Sb drives the southeast--northwest outflow. T Tauri Sb has recently faded probably because of increased extinction as it passes through the southern circumbinary disk. While Sb is approaching periastron, T Tauri Sa instead has brightened and is detected in all our \textit{J}-band imagery for the first time.}
   {}

   \maketitle
%

\section{Introduction}

T Tauri is the historical prototype of young and accreting low-mass stars. It is located in the Taurus-Auriga star-forming region at a distance of $146.7\pm0.6$\,pc \citep{loinard07} with an age of 1-2 Myr \citep{kenyon95}. At optical wavelengths, T Tauri N is an early K star \citep{cohen79}. From stellar evolutionary models and the effective temperature in the optical, \citet{schaefer20} estimated T Tauri N's mass to be ${\sim}2.1M_\odot$. The recent analysis of high-resolution near-infrared (NIR) spectra, however, suggests a significantly ($\approx$ 1000 K) lower photospheric temperature of T Tauri N than the temperature measured at optical wavelengths and also a surprisingly low surface gravity \citep{flores20}. It is therefore possible that T Tauri N is younger and of a lower mass than previously thought. 

Using NIR speckle interferometry, \citet{dyck82} found a companion at 0$\farcs$7 separation to the south of T Tauri N. This infrared (IR) companion, T Tauri S, is very red, and its brightness fluctuates at all NIR and mid-infrared (MIR) wavelengths \citep{ghez91, beck04}. Using speckle holography, \citet{koresko00} found that T Tauri S is itself a close (${\sim}0\farcs1$) binary, composed of the IR luminous T Tauri Sa and the early M-star companion, T Tauri Sb. The masses of T Tauri Sa and Sb are  ${\sim}2.1M_\odot$ and ${\sim}0.48M_\odot$, respectively. The orbit of the T Tauri Sa-Sb binary has a period of $\sim$27 years, a semi-major axis of $\sim$12.5~AU, and an inclination of $\sim$20~degrees to the plane of the sky \citep{koehler15,schaefer14,schaefer20}. The mass of T Tauri N is only loosely constrained by the current orbital solutions. 

The extinction toward T Tauri N was estimated by \citet{kenyon95} to be $A_V=1.39$~mag. T Tauri Sa appears to be heavily extinct and supposedly surrounded by an almost edge-on, small (3-5~AU radius), circumstellar disk hiding the stellar photosphere \citep[e.g.,][]{koresko00,kasper02,beck04,duchene05}. T Tauri Sa is variable in the NIR to the MIR on short and longer timescales most likely by processes involving variable accretion and extinction \citep{vanBoekel10}. The extinction toward T Tauri Sb is also much higher than the one toward T Tauri N and changed only moderately around an average value of $A_V=15$~mag \citep{duchene05} for a large part of its orbit. Since 2015, T Tauri Sb has faded significantly in the NIR \citep{schaefer20}. It is likely that the recently detected, roughly north--south oriented, and highly inclined southern circumbinary disk \citep{yang18, manara19} obscures T Tauri Sb while it passes through the disk plane on its orbit \citep{koehler20}.

The MIR interferometric observations by \citet{ratzka09} resolved the circumstellar disk around T Tauri Sa and derived that it is inclined by 72 degrees and roughly oriented north--south on the sky. The circumstellar disk around T Tauri Sb instead might be approximately coplanar with the southern binary's orbit with significant uncertainty. T Tauri N and its circumstellar disk are instead seen not far from pole-on with inclinations to the line of sight between $8\degree-13\degree$ \citep{herbst86} and $28\degree$ \citep{manara19}.

The T Tauri stars also drive at least two outflow systems with sky projected position angles, which are nearly perpendicular to each other \citep{boehm94, herbst96}. One outflow has a relatively high inclination of ${\sim}70\degree$ to the line of sight \citep{gustafsson10, kasper16} in the southeast--northwest direction, and another one in the northeast--southwest direction has radial velocities exceeding 100 km~s$^{-1}$ \citep{boehm94} with a low inclination to the line of sight of $23\degree$ \citep{eisloeffel98}. The near environment of T Tauri is also a source of surprisingly strong emission of molecular hydrogen. The spatial distribution of the H$_2$ emission displays a complex pattern tracing the two main outflows at all observed angular scales \citep{herbst97, herbst07, saucedo03, gustafsson10}. Most of the H$_2$ emission is likely to be generated by shock heating \citep{herbst97}, but \citet{saucedo03} also found ultraviolet (UV)-fluorescent H$_2$ emission, suggesting the action of low density, wide opening-angle outflows driving cavities into the molecular medium.

In this paper, we present new NIR broad-band and H$_2$ high-contrast imaging of the T Tauri triple system with VLT-SPHERE and LBT-LUCI observed between October 2016 and December 2018. We resolve the inner system at the diffraction limit of the 8-m class telescopes and report on a number of new spatial features, which help us to better understand the complex system. By comparing the new imagery to our early SPHERE observations obtained in December 2014 during a science demonstration \citep{kasper16}, we trace the tangential motion of several of the spatial features and are able to assign those to the main outflows. We further present \textit{J}-, \textit{H}-, and \textit{K}-band photometry of the stars and concur with \citet{koehler20} that we are currently seeing T Tauri Sb passing through the southern circumbinary disk. We finally report the detection of T Tauri Sa in the \textit{J}-band for the first time, which presents an opportunity for follow-up spectroscopy of photospheric emission and the determination of its spectral type. 

\section{Observations and data reduction}

\subsection{VLT-SPHERE}
We observed T Tauri with VLT-SPHERE \citep{beuzit19} on various occasions between 2016 and 2018 in the frame of SPHERE Guaranteed Time Observations (GTO). Details about the individual observations are summarized in Table \ref{tab-sphere_obs}. We also obtained LBT-LUCI data as part of its adaptive optics (AO)-commissioning on 22 October 2016 and 23 November 2016; the details of which are  provided in Table \ref{tab-luci_obs}.

The SPHERE data were recorded with the apodized Lyot coronagraph \citep{carbillet11} with an inner working angle close to 100 mas to suppress the point spread function (PSF) of the brighter T Tauri N, which also served as the guide star for the extreme AO system SAXO \citep{fusco14}. SPHERE employs the InfraRed Dual-band Imager and Spectrograph \citep[IRDIS,][]{dohlen08} and a low spectral resolution (R$\sim$30) Integral Field Spectrograph \citep[IFS,][]{claudi08} for the \textit{J}- and \textit{H}-bands. The observations were carried out in pupil-stabilized mode, but only the December 2017 observations were  long enough to cover sufficient field rotation for effective image processing with angular differential imaging (ADI) techniques \citep{marois06}. SPHERE observing sequences typically interleave the science exposures with so-called FLUX measurements, that is to say with an AO guide star that is moved off the coronagraph and a neutral density (ND) filter that is in the beam to avoid detector saturation. These FLUX exposures can be used for relative photometry. 

We used the SPHERE data reduction pipeline \citep{pavlov08} to create backgrounds, bad pixel maps, and flat fields. We reduced the raw data by subtracting the background, replacing bad pixels by the median of the nearest valid pixels, and finally by dividing the images by the flat field. We also used the pipeline to create the IFS x-y-$\lambda$ data cube. Parts of this cube were collapsed along the wavelength axis to create broad-band images in the \textit{J}-band (1140-1350~nm) and the short end of the \textit{H}-band (1490-1640~nm). We note that the long wavelength end of the \textit{H}-band is cut off by the IFS band-selection filter. The November 2017 and January 2018 IRDIS and IFS data were processed by classical ADI, that is, by subtracting the median of all images before de-rotation and averaging. The gentle processing is necessary because a more aggressive ADI processing may produce significant artifacts, for example, via the principle component analysis \citep[e.g.,][]{amara12} of a complex object with a wealth of azimuthally distributed spatial structure similar to T Tauri. No ADI techniques were applied to the relatively short 2016 and 2018 data sets due to insufficient field rotation during the observation. We applied the IRDIS and IFS plate scales of $12.255\pm0.021$\,mas and $7.46\pm0.02$\,mas per pixel, respectively, and the true north orientation is the same as the one provided in the SPHERE user manual \citep[see also][]{maire16}. 

To obtain precise photometry, the unsaturated PSF of T Tauri N from the FLUX measurements was corrected for the attenuation by the ND filter\footnote{see https://www.eso.org/sci/facilities/paranal/instruments/sphere/inst/filters.html} and used to simultaneously fit the T Tauri Sa and Sb binary after local background subtraction through minimization of the residuals. This data reduction strategy is vastly superior to simple aperture photometry in crowded areas with PSF overlap, such as T Tauri. We estimated the errors of this procedure by introducing small random offsets in the local background (3$\sigma$) and the assumed positions ($<$1 pixel) of the stars. 

\begin{table*}
\caption{VLT-SPHERE observations of T~Tauri.}          
\label{tab-sphere_obs}     
\centering                       
\begin{tabular}{l l l l l}       
\hline\hline                
Date & Filter & Central wavelength & Exposure time & Comments \\    
\hline
    18 Nov 2016 & B-\textit{H} (IRDIS)  & 1.625~\mum & 576 s          & Prog-ID 198.C-0209(N) \\
    30 Nov 2017 & \textit{J} (IFS)      & 1.250~\mum & 4032 s         & Prog-ID 1100.C-0481(C) \\
                & \textit{H} (IFS)      & 1.575~\mum & 4032 s         &   \\
                & \textit{K}1 (IRDIS)   & 2.110~\mum & 3808 s         &   \\
    06 Jan 2018 & \textit{J} (IFS)      &            & 5376 s         & Prog-ID 1100.C-0481(C) \\
                & \textit{H} (IFS)      &            & 5376 s         &   \\
                & \textit{K}1 (IRDIS)   &            & 5078 s         &   \\
    14 Dec 2018 & B-\textit{J} (IRDIS)  & 1.245~\mum & 512 s          & Prog-ID 1100.C-0481(K), NIR ND1 \\
                & B-\textit{H} (IRDIS)  &            & 384 s          &   \\
                & B-\textit{K$_s$} (IRDIS) & 2.182~\mum & 128 s          &   \\
\hline                                  
\end{tabular}
\end{table*}

\subsection{LBT-LUCI}
LUCI \citep{seifert03,heidt18} is the facility NIR (0.95~\mum{} to 2.4~\mum) imager and spectrograph at the LBT. LUCI can work in seeing-limited mode with a 4\arcmin~FoV, as well as in AO imaging and long-slit spectroscopy mode with a 30\arcsec~FoV. LUCI uses LBT's First Light Adaptive Optics (FLAO) system \citep{esposito12} to correct atmospheric turbulence. 

T Tauri was observed with LUCI as part of its AO-commissioning on 22 October 2016 and 23 November 2016. Again, T Tauri N was used as the AO guide star. The observations were carried out in field-stabilized mode. Diffraction-limited imaging was acquired for the broad \textit{J}-band filter, as well as the H$_2$ and Br$\gamma$ narrow-band filters. Details of the observations are listed in Table~\ref{tab-luci_obs}. The standard set of daily calibration data were recorded.

\begin{table*}
\caption{LBT-LUCI observation details on T~Tauri.}          
\label{tab-luci_obs}     
\centering                       
\begin{tabular}{l l l l}       
\hline\hline                
Parameter & \textit{J}-band & H$_2$ & Br$\gamma$ \\    
\hline 
                     
   Wavelength ($\lambda_c\pm\Delta\lambda$)& 1250$\pm$150 nm & 2127$\pm$12 nm & 2171$\pm$12 nm \\
   Date of observation & 23 November 2016 & 22 October 2016 & 23 November 2016 \\
   Exposure time        & 2.6 s          & 2.6 s & 2.6 s \\
   Co-adds                      & 24                    & 24 & 24  \\
   Number of frames     & 10                    & 18 & 24 \\
   Total exposure time& 10.40 minutes  & 18.72 minutes  & 24.96 minutes \\
   Seeing conditions    & 0\farcs37 - 0\farcs95  & 0\farcs50 - 1\farcs00  & 0\farcs37 - 0\farcs95 \\
   Saturated            & T Tauri N       & T Tauri N, T Tauri Sa  & T Tauri N, T Tauri Sa \\
\hline                                  
\end{tabular}
\end{table*}

In order to mitigate LUCI's significant detector artifacts, we dithered the field of view between two positions A and B in an A-B-B-A pattern to subtract the sky, bias, and dark current simultaneously. The dithering, however, leaves channel cross-talk residuals that are easily visible in the LUCI images as horizontal circular structures. The effect is more pronounced in the \textit{K}-band frames (Br$\gamma$ and H$_2$), where both T Tauri N and T Tauri Sa are saturated. 


We followed standard procedures for LUCI data reduction using IRAF. As a first step, bad pixel maps and master flat fields were created using the calibration files and then applied independently to each of the science frames. To subtract the sky, bias, and dark current, dithered pairs were subtracted, which in turn created a positive image on one quadrant and a negative image on the other. For these positive and negative images, mask images were created at known artifact locations. The science images were then multiplied by the corresponding mask images. The resulting frames were then centered on T Tauri Sa (\textit{J}-band) or Sb (H$_2$ and Br$\gamma$). The H$_2$ line emission map was produced by removing the nearby continuum from the H$_2$ image, which were both normalized to T Tauri Sb. We used the Br$\gamma$ as a continuum filter. While T Tauri Sb shows spatially unresolved Br$\gamma$ emission with an equivalent width of $\sim$0.6~nm \citep{duchene05}, the resulting error in the continuum flux of $\sim$2.5\% for the 24~nm FWHM LUCI filters is negligible compared to other normalization uncertainties such as variations in the color of the stars and the circumstellar material between the two filters. The many radial lines in the final images in Figures~\ref{fig:luci_J1} and \ref{fig:luci_H2line} were produced by the rotating diffraction pattern from the spiders holding the adaptive secondary and tertiary of the LBT in our field-stabilized observations. 

\section{Results}
\subsection{Continuum imaging}

\begin{figure*}[ht]
   \centering
   \includegraphics[width=\hsize]{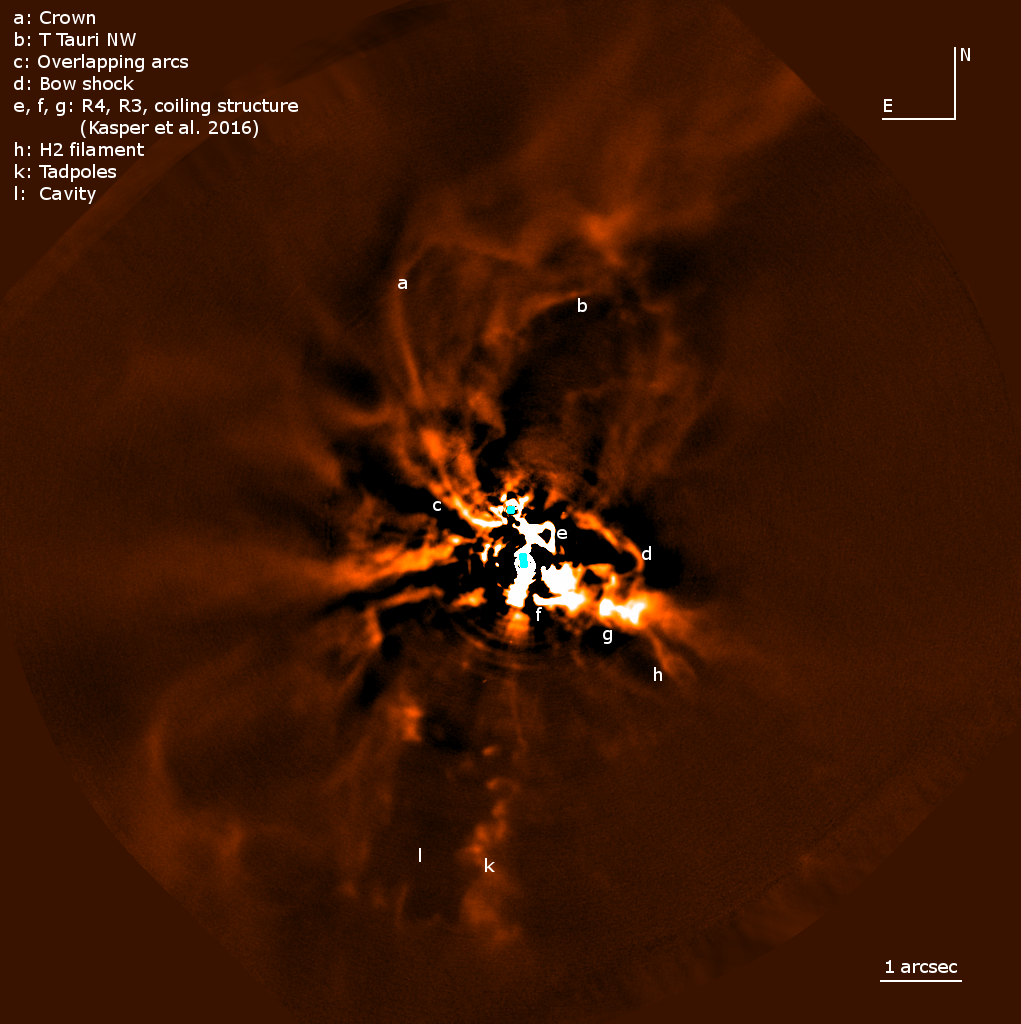}
      \caption{SPHERE-IRDIS classical ADI image of T Tauri in the \textit{K}1-filter in linear scale. The image combines data recorded in November 2017 and January 2018. New features and some known ones (such as T Tauri NW) are labeled in the image. The stars are either suppressed by the coronagraph (T Tauri N) or saturated by the color-scale (T Tauri Sa and Sb); their positions are indicated by the cyan dots. The dark patches are artifacts from the ADI processing. The main spatial structures discussed in this paper are indicated.}
         \label{fig:K1_ADI_Dec2017}
   \end{figure*}

Figure~\ref{fig:K1_ADI_Dec2017} shows the T Tauri system observed in the SPHERE/IRDIS dual-band imaging \citep[DBI,][]{vigan10} \textit{K}1 ($\lambda_\mathrm{c} = 2110$~nm) filter. This image was ADI processed by subtracting the median of the images before de-rotating and averaging them. It shows several previously unknown features as well as all the structures R1-4 and the coil, which were previously identified \citep{kasper16}. Some of the features labeled in Figure~\ref{fig:K1_ADI_Dec2017} are new, including the following.
\begin{description}
\item[\textbf{"a":}] A large scale roughly circular structure of 3\farcs5 diameter just north of T Tauri encompassing the H$_2$ region T Tauri NW. This structure has two bump-shaped extensions to the north, which are reminiscent of a crown.  
\item[\textbf{"c":}] Some overlapping arcs extending from just south of T Tauri N to the northeast. 
\item[\textbf{"d":}] A double bow shock about 1-2 arcseconds west of T Tauri, which also emits in H$_2$ at 2.122~$\mu$m as shown in Figure~\ref{fig:luci_H2line}.
\item[\textbf{"h":}] A small H$_2$ filament to the southwest, which appears to be connected to the coil. This feature can be seen in 2.122~$\mu$m H$_2$ emission, which is shown in Figure~\ref{fig:luci_H2line} as well.
\item[\textbf{"k":}] A wiggling structure with individual clumps, which we labeled as tadpoles because of its appearance. The tadpoles start at roughly 3\arcsec{} south of T Tauri N and extend to the south.
\end{description}

\begin{figure}
   \centering
   \includegraphics[width=\hsize]{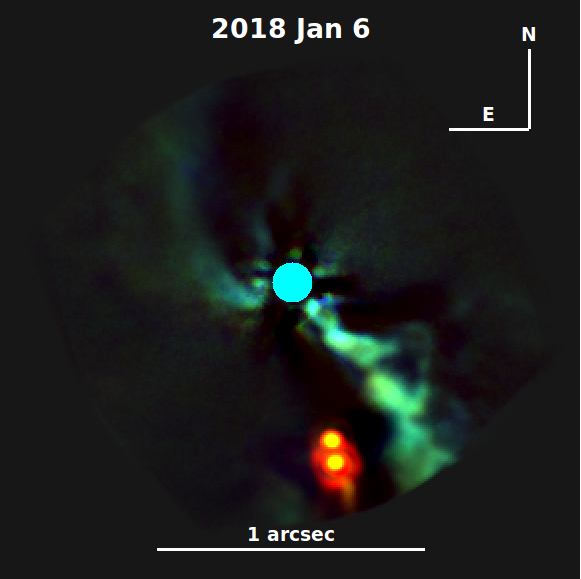}
      \caption{SPHERE-IFS classical ADI \textit{YJH}-composite image of T Tauri in linear scale recorded in January 2018.}
         \label{fig:J_IFS_Dec2017}
   \end{figure}

The overlapping arcs extending from just south of T Tauri N to the northeast are also detected in the simultaneously recorded IFS \textit{YJH}-composite image shown in Figure~\ref{fig:J_IFS_Dec2017}, which also shows a clear detection of T Tauri Sa and Sb. The picture also shows the comma-like structure, previously labeled R2 \citep{kasper16}, extending from T Tauri Sa to the south. It also shows the R4 feature to the southwest of T Tauri N labeled "e" in Figure~\ref{fig:K1_ADI_Dec2017}. All of the features seen in the IFS image appear in all of its wavelength channels across the NIR and hence they represent reflection nebulosity structures. No significant flux enhancement was observed in the IFS channels covering prominent emission line features, such as Pa$\beta$ or Fe II, at the R$\approx$30 spectral resolution.

\begin{figure*}
   \centering
   \includegraphics[width=\hsize]{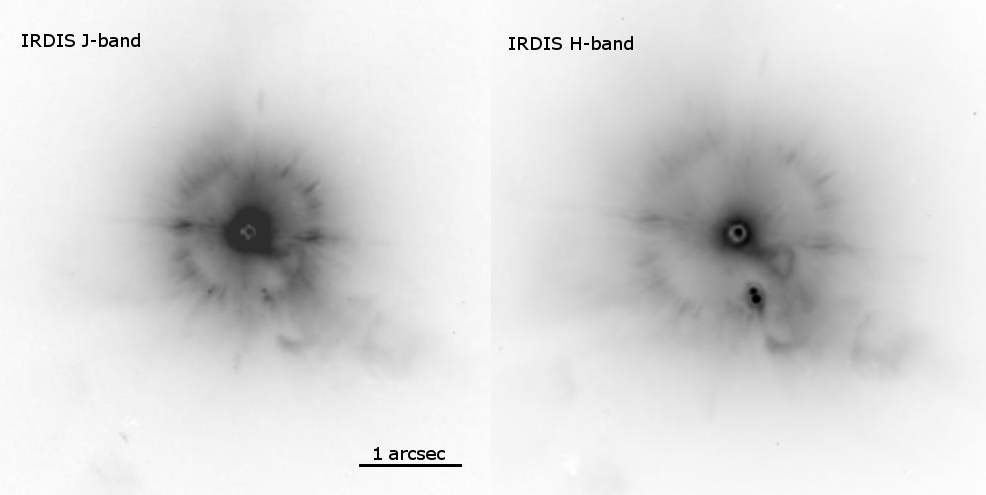}
      \caption{IRDIS coronagraphic images of T Tauri in \textit{J}-band (left) and \textit{H}-band (right). These images were recorded in December 2018. No ADI processing has been applied to these relatively short exposures.}
         \label{fig:JvsH_Dec2018}
   \end{figure*}

Non-ADI coronagraphic images of T Tauri in the \textit{J}- and \textit{H}-bands are displayed in Figure~\ref{fig:JvsH_Dec2018}. The structures that scale in radial distance to the center with wavelength, are PSF artifacts. For example, the prominent ring of elongated speckles belongs to the AO coronagraphic PSF and it indicates the so-called AO control radius \citep{perrin03}, that is, the radial distance up to which the AO system can clean the PSF. Structures that do not scale with wavelength are real, such as the reflection nebulosity features R3, R4, and the coil. T Tauri Sa is detected again in the \textit{J}-band imagery of December 2018.

\begin{figure}
   \centering
   \includegraphics[width=\hsize]{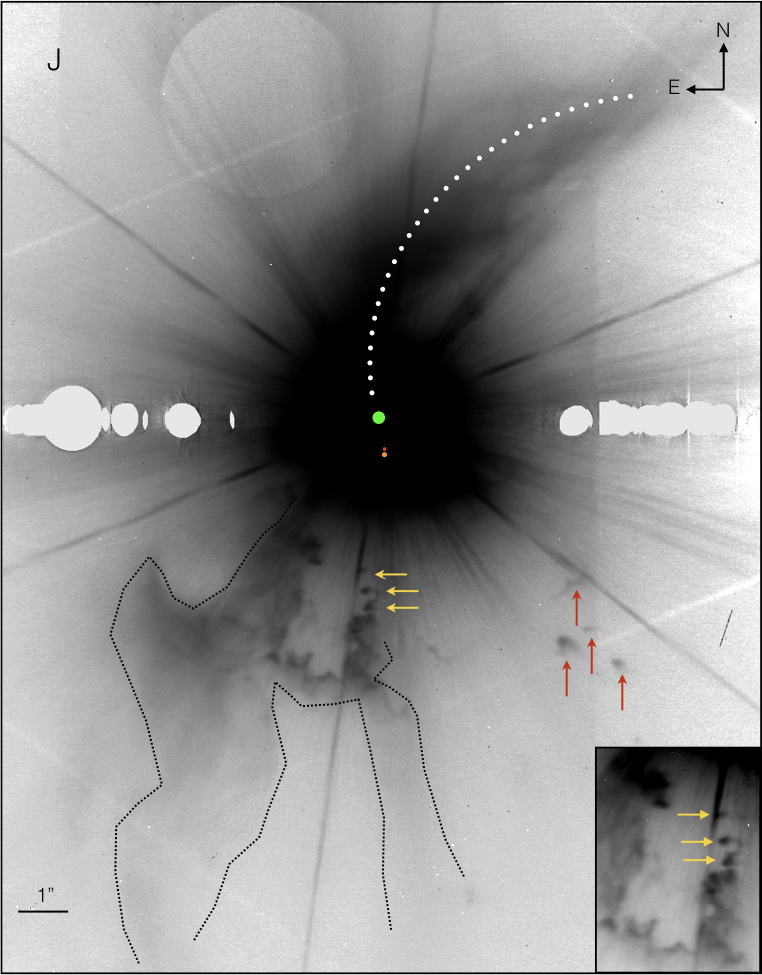}
      \caption{LUCI \textit{J}-band image of T Tauri from November 2016 displayed on a logarithmic scale. The dotted arc indicates the northern arm of the reflection nebulosity discussed by \cite{stapelfeldt98}. The yellow arrows indicate the tadpoles. The red arrows mark bow-shaped features. Extended features around an apparent cavity are enclosed within the black dotted lines. The green, orange, and red dots identify the locations of T Tauri N, T Tauri Sa, and T Tauri Sb, respectively. The inset shows the tadpoles and the cavity with a different gray-scale cut.}
         \label{fig:luci_J1}
   \end{figure}

Figure \ref{fig:luci_J1} shows the LUCI \textit{J}-band image of T Tauri observed in November 2016. The yellow arrows indicate the tadpole structure labeled "k" in Figure~\ref{fig:K1_ADI_Dec2017}. New extended features are detected in the south--southeast direction. A cavity pointing toward the stars also appears distinctly and falls on a line connecting it to the H$_2$ emitting region T Tauri NW. Other new features include the knots, linear filament, and bow-shaped features in the \textit{J}-band image toward the southwest of the stars, which can also be seen in Figure \ref{fig:luci_J1}. All these features lie roughly in the same direction as the coiling structure, but they are further away. 

In the north of T Tauri, we see the northern part of the arc-like reflection nebulosity, which was first detected by \citet{stapelfeldt98}. While most structures to the south of T Tauri have a similar appearance at all NIR wavelengths, this arc is much brighter at shorter wavelengths. The crown-shaped feature labeled "a" in the \textit{K}-band image in Figure~\ref{fig:K1_ADI_Dec2017} is not visible in the \textit{J}-band.

\begin{figure}
  \centering
  \includegraphics[width=\hsize]{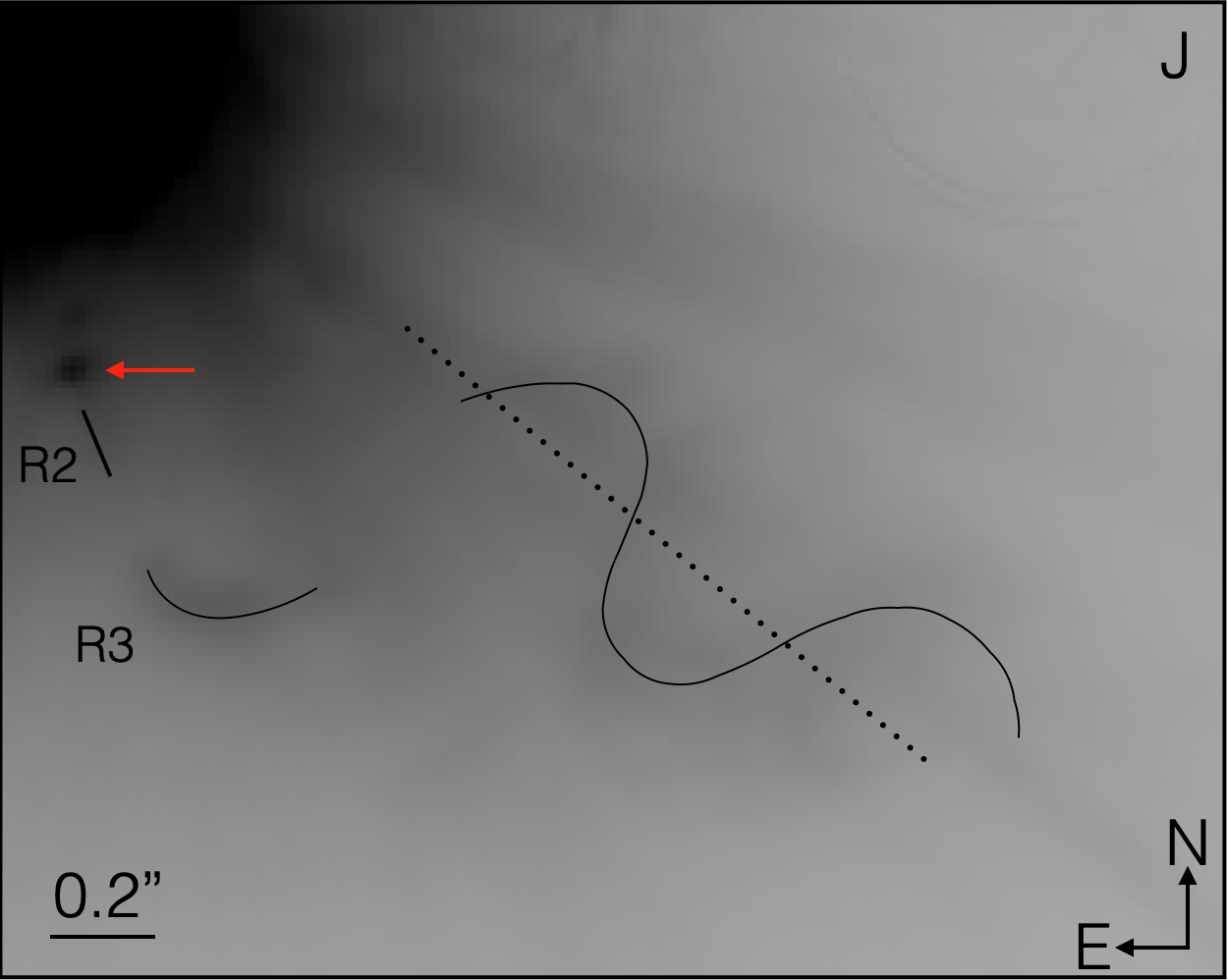}
      \caption{Zoom of Figure~\ref{fig:luci_J1} on the area southwest of T Tauri N showing the coiling structure and the reflection nebulosity features R2 and R3 (as shown in Figure 2 of \cite{kasper16}) in the vicinity of T Tauri S. The dotted line connects the inflection points of the coil. The red arrow indicates T Tauri Sa.}
         \label{fig:luci_J2}
  \end{figure}

Figure \ref{fig:luci_J2} shows the central area of the November 2016 LUCI \textit{J}-band image at a different scaling, revealing the first-ever detection of T Tauri Sa in the \textit{J}-band. T Tauri Sa was even quite a bit brighter than T Tauri Sb at that time. Before 2016, the faintness of the photometrically variable star and the sensitivity of the available instrumentation had only allowed us to detect T Tauri Sa at \textit{H}-band or longer wavelengths. Despite the high spatial resolution of the LBT in the \textit{J}-band of about 30~mas corresponding to 4.3~AU, the image of T Tauri Sa still appears unresolved, and we do not see evidence for spatially resolved emission scattered from T Tauri Sa's circumstellar disk. 


\subsection{Molecular hydrogen emission}

\begin{figure}
   \centering
   \includegraphics[width=\hsize]{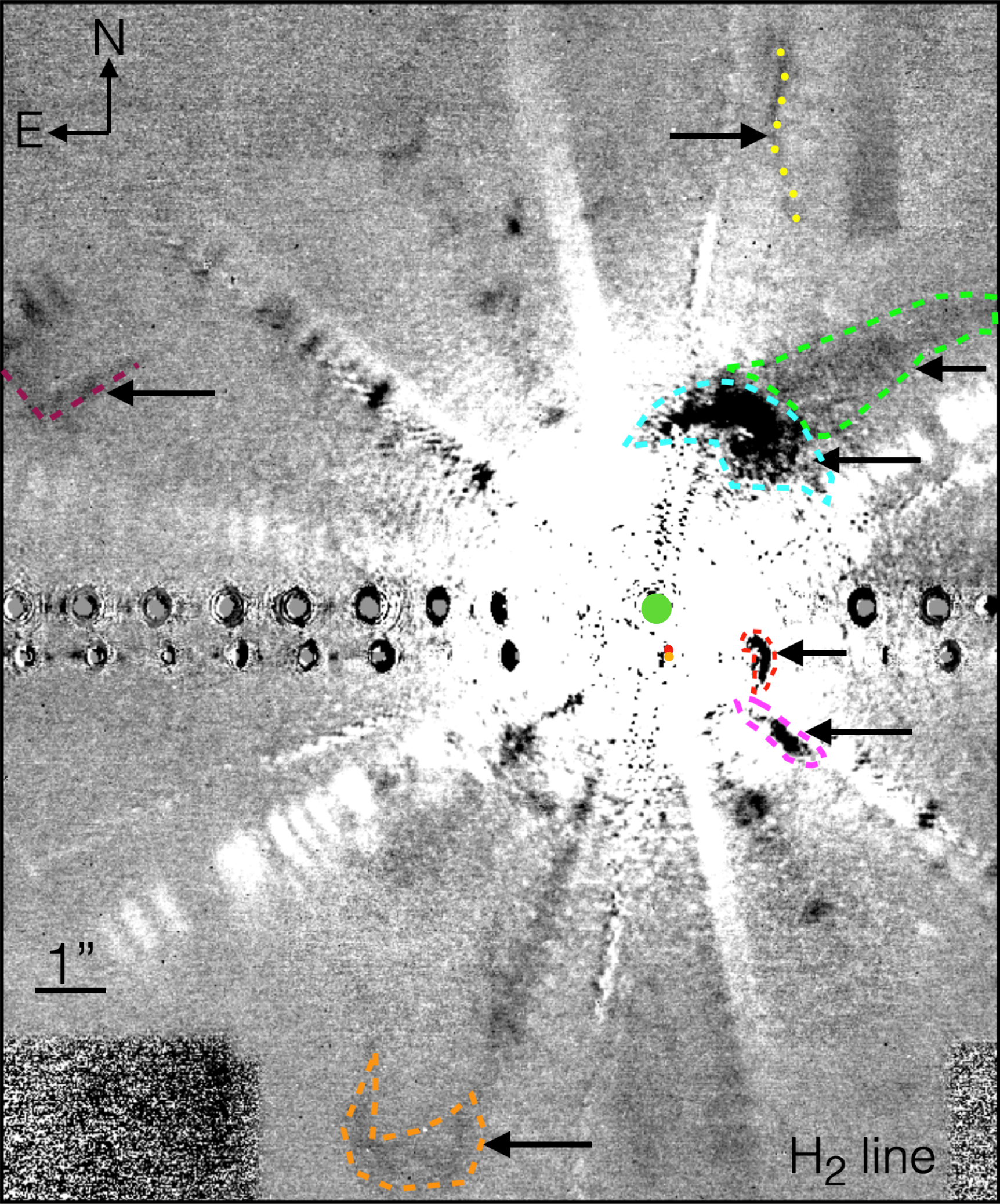}
      \caption{T Tauri H$_2$ line emission map obtained with LBT-LUCI in October 2016. Only features which cannot be explained by image processing artifacts are indicated and discussed in the text. The positions of the stars are indicated by colored dots (T Tauri N, Sa, and Sb are in green, orange, and red, respectively).}
         \label{fig:luci_H2line}
   \end{figure}

The T Tauri H$_2$ line emission map is shown in Figure \ref{fig:luci_H2line} and was obtained by subtracting the Br$\gamma$ image (as a continuum) from the H$_2$ image. The map suffers from significant residuals. Firstly, the H$_2$ and Br$\gamma$ observations were collected with a time gap of one month and at different hour angles. So, the diffraction spikes produced by the telescope's secondary mirror support structure are located at different position angles, and instrument internal aberrations differ. Secondly, the radial distance from the center of all PSF residuals scales with wavelength, which hampers the effectiveness of the subtraction using  narrow-band filters whose central wavelengths differ by 44~nm. Nevertheless, there are several H$_2$ emission features that cannot be attributed to image reduction artifacts. 

In addition to the already known T Tauri NW \citep{herbst96}, the bow shock feature reported by \citet{herbst07}, and the southwestern filament \citep{kasper16}, which can all be seen in Figure~\ref{fig:K1_ADI_Dec2017}, we detected several new features as follows.\
\begin{description}
\item[\textbf{green:}] A relatively faint (compared to T Tauri NW) extended feature, which appears to be extending from T Tauri NW in the northwest direction. 
\item[\textbf{yellow:}] A curvy linear feature toward the northwest of the image.
\item[\textbf{maroon:}] An L-shaped feature toward the east of the image.
\item[\textbf{orange:}] A low signal-to-noise ratio (S/N) patch of H$_2$ line emission that is far away from the stars in the south--southeast direction.
\end{description}

\subsection{Tangential motion of circumstellar structures}

Some of the reflection nebulosity structures in the T Tauri system, such as R3, R4, and the coil, show a significant tangential, that is to say sky projected motion between the high-spatial resolution \textit{J}-band images of December 2014 \citep{kasper16} and December 2018. While, in principle, a change in the underlying illumination pattern could also mimic tangential motion, the motion of the southern stars, which illuminate R3, R4, and the coil \citep{yang18}, relative to the features is relatively minor over the time span of our observations. So a significant change in the illumination pattern due to stellar motion is unlikely. Shadowing effects due to circumstellar material blocking the line of sight between a star and a feature \citep[e.g.,][]{keppler20} could also change the appearance of reflection nebulosity structures on short timescales. Shadowing is, however, unlikely to mimic the approximately rigid motion of the features we observe in T Tauri. Still, we cannot exclude such effects at this point in time, but continued monitoring of the features will eventually allow us to unambiguously confirm tangential motion.

\begin{figure}
   \centering
   \includegraphics[width=\hsize]{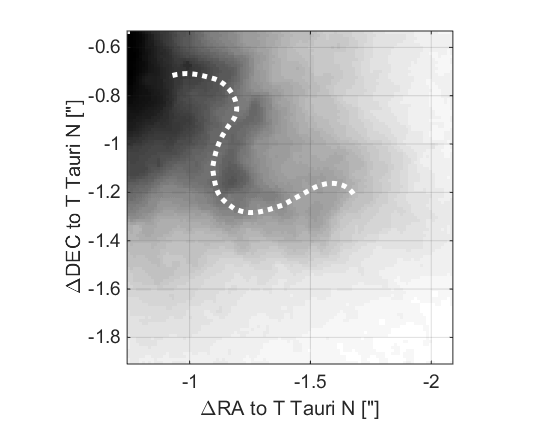}
      \caption{IRDIS\textit{J}-band image of the coiling structure taken in December 2018. The white dotted line indicates the coiling structure's position in December 2014 \citep[see][]{kasper16}.}
         \label{fig:coil2014vs2018}
   \end{figure}

\begin{figure}
   \centering
   \includegraphics[width=\hsize]{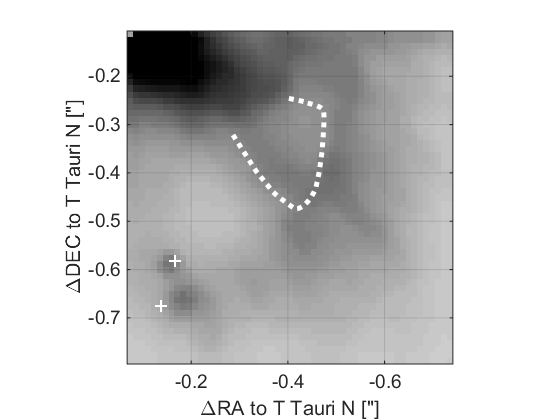}
      \caption{IRDIS \textit{J}-band image of the R4 structure taken in December 2018. The white dotted line indicates R4's position, and the "+" symbols indicate the positions of T Tauri Sa and Sb in December 2014 \citep[see][]{kasper16}.}
         \label{fig:r42014vs2018}
   \end{figure}

\begin{figure}
   \centering
   \includegraphics[width=\hsize]{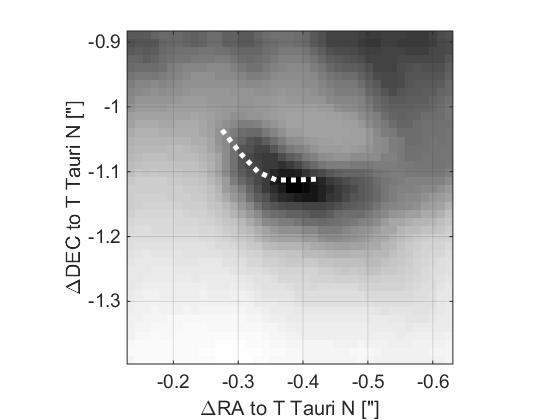}
      \caption{IRDIS \textit{J}-band image of the R3 structure taken in December 2018. The white dotted line indicates R3's position in December 2014 \citep[see][]{kasper16}.}
         \label{fig:r32014vs2018}
   \end{figure}

Here, we assume that we are indeed observing tangential motion, which allows us to associate structures in outflow systems with the stars from which they originate. The Figures~\ref{fig:coil2014vs2018} to \ref{fig:r32014vs2018} use T Tauri N as the center field. The motion of stars with respect to each other and the reference frame must be considered to interpret the observed structure dynamics. Over our observation period, T Tauri Sa and Sb moved relative to to T Tauri N by roughly 10~mas yr$^{-1}$ to the west-northwest and about 5~mas yr$^{-1}$ to the southeast, respectively \citep{koehler15}. The change in position of the two stars can also be observed in Figure~\ref{fig:r42014vs2018}. 

Figure~\ref{fig:coil2014vs2018} shows, for example, that the coiling structure moved by about 50-100~mas to the west--southwest in four years, which would correspond to 1.75-3.5~AU~yr$^{-1}$ or 8-16~km~s$^{-1}$ at the distance of T Tauri. It is not possible to give a more precise value because the coiling structure's morphology also slightly changed over this period. The velocity is quite comparable to the tangential motion of T Tauri NW \citep{kasper16}.

For the following discussion, we follow the structure designations R3 and R4 defined by \citet{kasper16}. The R4 structure shown in Figure~\ref{fig:r42014vs2018} moved significantly to the northwest by about 100-200~mas (3.5-7~AU~yr$^{-1}$ or 16-32~km~s$^{-1}$). Despite its sky-projected proximity of $\sim$0\farcs4 (60~AU) to T Tauri N, the motion in an azimuthal direction at a velocity exceeding Keplerian velocity ($\sim$5--6~km~s$^{-1}$ for the $\sim{}2.1M_\odot$ star) by a factor of a few excludes an association with this star. Instead, the R4 structure appears to be part of the northwest outflow driven by T Tauri S, a conclusion that has already been derived from polarimetric data showing that R4 is illuminated by T~Tauri~S \citep{yang18}.

The R3 structure shown in Figure~\ref{fig:r32014vs2018} exhibits an on-sky motion of 30-40 mas mostly to the west and slightly south, corresponding to about 1~AU~yr$^{-1}$ or 5~km~s$^{-1}$. The spatial location that is not far from the coil and the similar direction of motion suggests that R3 is associated to the same northeast--southwest outflow system and probably represents swept-up material in the periphery of the outflow.

After the clear detection of the tadpole structure in the SPHERE (in Figure~\ref{fig:K1_ADI_Dec2017}) and LUCI (in Figure~\ref{fig:luci_J1}) data, we reanalyzed the 2014 \textit{J}-band data presented by \citet{kasper16} and were also able to detect the tadpoles in these data at a relatively low S/N. The tadpole structure, however, appears to be relatively stationary. No apparent motion in the plane of the sky was measured over the period of our observations between 2014 and 2017. 

\section{Discussion}

\subsection{New features in the system}

 The pillar-like structures to the southeast of T Tauri are the most evident on a larger field of view, with the tadpoles being the most prominent structures. There is a well-defined cavity to the east of the tadpoles, which terminates in a box-like structure at the south end. There is an area with H$_2$ line emission further away in the same direction (in Figure~\ref{fig:luci_H2line}). The cavity and tadpoles are oriented such that they connect on a straight line through T Tauri S to T Tauri NW. We consider this as more evidence, in addition to the tangential motion of T Tauri NW \citep{kasper16}, that we see a southeast--northwest oriented bipolar outflow system driven by T Tauri S. Further support for this model comes from the relatively low radial velocities of the southeast--northwest outflow reported by \citet{boehm94} and \citet{herbst97}, which is hardly compatible with an outflow from the nearly face-on T Tauri N.

The knotted wiggling structure of the tadpoles is reminiscent of jets seen in several young stellar objects typically traveling at several tens to hundreds km s$^{-1}$ \citep{bally16}. However, their stationary appearance over a period of four years with respect to the stars and their location at a rather sharply defined western edge of a cavity suggest a different explanation. The tadpoles could represent swept-up material at the edge of the outflow cavity showing only a modest level of entrainment.

Another intriguing feature is the narrow, roughly circular structure, which we label "crown" in Figure~\ref{fig:K1_ADI_Dec2017}. This structure is not seen at wavelengths shorter than the \textit{K}-band, where it is replaced by the much more fuzzy arc of reflection nebulosity found by \citet{stapelfeldt98}. The strikingly variable appearance between the different NIR observing bands can be reconciled by the crown being a wind-inflated bubble \citep{herbst97} of material swept up by the southeast--northwest outflow, which is seen through Stapelfeld's arc associated with T Tauri N. This is consistent with a system model where the close to face-on T Tauri N lies in front of T Tauri S, which launches the outflow pointing away from the observer in its northwestern part. At shorter wavelengths, the arc is brighter due to the increased scattered light from T Tauri N \citep{yang18}, and at the same time, the crown becomes more highly extinct and fainter.

The relative distance to the T Tauri stars can be determined from the orbital solution and radial velocity (RV) measurements. According to the best-fit T Tauri N-S orbit \citep{schaefer20}, T Tauri S went through the line of nodes in 1936 with an RV difference to T Tauri N of about $\pm$6.5~km~s$^{-1}$. The sign of the RV difference cannot be determined from an astrometric orbit solution. \citet{duchene05} found that the heliocentric RVs of T Tauri Sa and Sb appear to be larger than the one of T Tauri N \citep{hartmann86} by 2-3~km~s$^{-1}$. These values are compatible with the orbit prediction within the error bars if T Tauri S previously went through the ascending node. Since then, T Tauri S has been moving behind T Tauri N for about 80 years and would now be located roughly 90~AU behind T Tauri N. However, the N-S orbit is still only loosely constrained and orbital solutions may exist for which T Tauri S is located in front of T Tauri N, as suggested by \citet{beck20}. 

Figure~\ref{fig:K1_ADI_Dec2017} also shows a bow shock at $\sim$1\farcs6 west of T Tauri. This bow shock is a signpost of the east--west outflow \citep{boehm94, herbst97} with the western part approaching (blue-shifted) at about 100~km~s$^{-1}$, which is consistent with an origin in the close to face-on T Tauri N. The bow shock also shows up in H$_2$ line emission in Figure~\ref{fig:luci_H2line}. The December 2002 image by \citet{herbst07} detected a bow shock labeled C1 about one arcsecond west of T Tauri S. Assuming that our bow shock and C1 are actually the same feature, it has moved by about 0\farcs5 to the west between December 2002 and December 2017, which translates into a tangential motion of about 4-5~AU~yr$^{-1}$ or 20-25~km~s$^{-1}$.  

The overlapping arcs extending from just south of T Tauri N to the northeast up to a distance of at least $\sim$0\farcs7 (100~AU) from the star can be seen in Figures~\ref{fig:K1_ADI_Dec2017} and \ref{fig:J_IFS_Dec2017}. Up to three individual arcs appear to fan out from a common location. The spatial extension of this structure is much larger than the radial size of T Tauri N's dust disk, which is estimated to lie between 0\farcs1 - 0\farcs15 (15-20~AU) \citep{manara19} and 43~AU \citep{akeson98}. Possible explanations for these structures include material from Stapelfeldt's arc streaming onto the disk. Such streamers have recently been observed at millimeter-wavelengths in class 0/I young stellar binaries \citep{alves19, pineda20}. Alternatively, the overlapping arcs could again represent swept-up materials in the periphery of an outflow cavity in the northeastern direction.

Finally, Figure~\ref{fig:luci_H2line} shows previously unknown regions with H$_2$ emission, which are aligned with the major outflows on roughly perpendicular axes that run southeast--northwest and northeast--southwest. We note that H$_2$ line emission arises in a number of physical processes, the most common of which are shock excitation and UV fluorescence. T Tauri is an interesting case in which both mechanisms occur \citep{vanLangevelde94,saucedo03,walter03}. Shock excitation is the dominant process at large distances to the star where the radiation from the star thins out. Jets and winds transfer momentum and entrain their  surroundings  by  means  of  shock  waves  propagating  into  the  medium.  These  shocks  tend to have much lower velocities than the jets themselves, typically a few tens of km s$^{-1}$, and they can be seen in H$_2$ emission when the interaction is with a molecular cloud. Often the emitting regions appear as bow shocks, similar to the well known T Tauri NW, the new weaker counterpart of T Tauri NW at about 9\arcsec{} to the south-southwest, and the compact bow shock to the west of T Tauri S. This bow shock is likely to be the one detected in 2002 and labeled C1 by \cite{herbst07}, and its on-sky motion is discussed above. We also see a long curvy feature north--northwest of the stars and an L-shaped feature toward the east of the image, which is almost in the same place as the E feature mentioned by \cite{herbst97}. 

\subsection{Photometric variability of the southern binary}

Both T Tauri Sa and Sb are photometrically variable. After a stable period since its discovery \citep{koresko00}, T Tauri Sb has started to dim in the \textit{K}-band since 2015 by now up to 2.5 magnitudes \citep{schaefer20}. At the same time, T Tauri Sa became brighter by about one magnitude on average with fluctuations of 0.5 - 1 magnitudes over the timescale of a few months. Given that the current orbital position of T Tauri Sb puts it at the location of the suspected T Tauri S circumbinary disk \citep{yang18, manara19}, the recent brightness fluctuations of T Tauri Sb are likely to originate from increased extinction toward T Tauri Sb while moving through the circumbinary disk plane on its orbit \citep{koehler20}.

Our photometry of T Tauri Sa and Sb is summarized in Table~\ref{tab:photometry} and support these results. In the absence of observations of a photometric calibrator, we assume that T Tauri N is not variable \citep{beck04} and has NIR magnitudes of $J = 7.1$, $H = 6.2$, and $K = 5.7$ \citep{herbst07}. The apparent magnitudes listed in the table were then computed from the observed delta magnitudes between T Tauri N, Sa, and Sb. The extinction toward T Tauri Sb during its photometrically moderately stable phase has been estimated to be A$_V \approx 15$~mag \citep{duchene05}. With a typical extinction ratio of $A_J / A_V = 0.26$ \citep[e.g.,][]{cieza05}, our observed additional extinction of A$_J \approx 3$~mag between December 2014 and December 2018 translates into an additional A$_V \approx 11.5$~mag resulting in a total extinction of A$_V \approx 27$~mag toward T Tauri Sb in December 2018. The additional extinction toward T Tauri Sb does not seem very large for a supposedly optically thick disk, but it is possible that Sb has not yet reached the disk's mid-plane. Further photometric monitoring of Sb could therefore allow us to perform tomography of the circumbinary disk.

We detected T Tauri Sa for the first time in the \textit{J}-band in November 2016 (see Figure~\ref{fig:luci_J2}), and we also detect it in our November 2017 and December 2018 data (Figures~\ref{fig:J_IFS_Dec2017} and \ref{fig:JvsH_Dec2018}), where it was of a similar brightness as T Tauri Sb. While a significant brightening of T Tauri Sa was also observed in the \textit{H}-band after 2014, its \textit{K}-band magnitude stayed rather constant considering the slightly variable central wavelengths of the filters used at different times and the very red spectrum of T Tauri Sa. This behavior cannot be explained by variable foreground extinction of a single emitter because extinction, albeit becoming smaller the longer the wavelength, affects all NIR wavelengths. It can, however, be explained by differential brightening of two flux emitting components, such as the star and its circumstellar disk. Compared to the stellar photosphere, excess emission from the disk is redder and often dominates the flux in the \textit{K}-band. The spectrum of T Tauri Sb, for example, shows a \textit{K}-band excess emission (also called spectral veiling) of $r_k \sim 2$, while no photospheric features could be detected for T Tauri Sa \citep{duchene05}. The \textit{K}-band flux of T Tauri Sa is therefore most likely dominated by disk emission. Compared to the \textit{K}-band, excess emission from T Tauri stars in the \textit{J}-band is smaller by factors of a few on average \citep{cieza05, edwards06}. Reduced extinction toward the photosphere of T Tauri Sa can, therefore, lead to a significant brightening in the \textit{J}- and \textit{H}-band where the star is a stronger contributor to the overall flux, while barely affecting the disk-dominated \textit{K}-band flux. This extinction must be produced locally, for example, by material close to the inner edge of the disk, for this mechanism to work. 

The \textit{J}-band detection of T Tauri Sa is particularly interesting because observations at this wavelength should suffer the least from spectral veiling. The \textit{J}-band maximizes the ratio between stellar photospheric emission and the combined excess from hot accretion shock emission (bright at short wavelengths, UV) and warm circumstellar dust or disk emission (bright at long wavelengths, \textit{K}-band to MIR) \citep{hartmann90}. Still \textit{J}-band veiling can be significant for T Tauri stars \citep{edwards06}, and the detection of photospheric features to determine the hitherto unknown spectral type of T Tauri Sa may still be hampered by shallow features to be detected in a low S/N spectrum.

\begin{table*}
\caption{
VLT-SPHERE NIR apparent magnitudes of T~Tauri Sa and Sb. The 2014 data were already published by \citet{kasper16} and based on IRDIS NB imaging and IFS. The central wavelengths for these data are 1.250, 1.575, and 2.217~\mum{} for \textit{J}, \textit{H}, and \textit{K}, respectively. The central wavelengths for the other observations slightly differ depending on the IRDIS filter and IFS, as given in Table~\ref{tab-sphere_obs}.
}          
\label{tab:photometry}     
\centering                       
\begin{tabular}{l l l l l l}       
\hline\hline                
Star & Filter & 9 Dec 2014 & 18 Nov 2016 & 30 Nov 2017 & 14 Dec 2018\\    
\hline
Sa & \textit{J} & $>17.5$ & & $15.52\pm0.06$ & $16.07\pm0.07$\\
   & \textit{H} & $12.35\pm0.2$ & $10.06\pm0.06$ & $11.14\pm0.04$ & $11.37\pm0.05$\\
   & \textit{K} & $7.9\pm0.1$ & & $8.14\pm0.08$ & $8.27\pm0.04$\\
Sb & \textit{J} & $13.3\pm0.1$ & & $15.28\pm0.07$ & $16.3\pm0.1$\\
   & \textit{H} & $10.8\pm0.05$ & $12.21\pm0.3$ & $12.17\pm0.04$ & $12.27\pm0.1$\\
   & \textit{K} & $8.9\pm0.1$ & & $10.5\pm0.4$ & $10.47\pm0.25$\\
\hline                                  
\end{tabular}
\end{table*}

\section{Conclusions}

In this paper, we report on the presence of various new spatial structures in the T Tauri system seen in NIR high-contrast imaging with VLT-SPHERE or seen in H$_2$ line emission maps obtained with LBT-LUCI. These new features naturally pose new challenges as to the interpretation of how they fit with our understanding of this enigmatic young triple star. It has long been known that there are at least two distinct, nearly perpendicular bipolar outflow systems in T Tauri. One is oriented in the southeast--northwest direction with the blue-shifted part approaching to the southeast, and another one is oriented east--west with the western part approaching at high velocity \citep{boehm94}. We measured the tangential motion of the feature R4, which is further compelling evidence for T Tauri S as the source of the southeast--northwest outflow \citep{kasper16}. On the other hand, \citet{ratzka09} were able to resolve the circumstellar disk around T Tauri Sa with MIR interferometry and deduced that it is seen close to edge-on with the disk plane oriented in the north-south direction. So T Tauri Sa's outflow should be roughly perpendicular to the disk axis, that is east-west, which leaves T Tauri Sb as the only possible origin for the southeast--northwest outflow. While T Tauri Sb is the least massive star in the system, it still emits significant Br$\gamma$ flux \citep{duchene05}. Therefore, Sb is undergoing active accretion, which may fuel the most prominent outflow in the system.

Structures to the west--southwest of T Tauri, such as the coil, the bow shock, and feature R3, were shown to move away from the stars mostly in the western direction at tangential velocities between 5 (R3) and 8-16 (coil) km s$^{-1}$. The bow shock instead moves at a somewhat higher tangential velocity of 20-25 km s$^{-1}$. It could well be that all of these features are part of a single outflow system driven by T Tauri Sa. This would be supported by the polarimetric observations reported by \citet{yang18}, which show that all structures in this direction appear to be illuminated by T Tauri S. An outflow from the close-to edge-on T Tauri Sa disk must, however, only be weakly inclined to the plane of the sky. It can therefore hardly launch the high radial velocity and likely highly inclined jet, which is observed in the east--west direction and attributed to T Tauri N \citep{boehm94}. We would, therefore, argue that both T Tauri N and Sa drive roughly northeast--southwest oriented outflows at different inclinations.

Another important observation is that T Tauri Sa has recently brightened enough to be detected with extreme AO instrumentation in the \textit{J}-band, which presents a great opportunity to detect photospheric features and determine the stellar spectral type. T Tauri Sb instead has faded, most likely by increased extinction while crossing the plane of the southern circumbinary disk on its orbit. Photometric monitoring of T Tauri Sb therefore presents an opportunity to perform tomography of this disk.  

As once written by \citet{menard01}, T Tauri is indeed the prototype for young stellar object complexity and once more showed to be a fascinating object, which demonstrates how ambient material swept up and entrained by misaligned outflows can create a multitude of spatial structure including bows, tadpoles, bubbles, and spirals patterns to name a few. It is a great example of how viewing geometry and general system complexity can lead to such a disarray, which can only slowly be disentangled by applying the whole arsenal of astronomical observation methods. 

\begin{acknowledgements}
This work is based on observations performed with VLT/SPHERE under program IDs 98.C-0209(N), 1100.C-0481(C) and 1100.C-0481(K). 

SPHERE is an instrument designed and built by a consortium consisting of IPAG (Grenoble, France), MPIA (Heidelberg, Germany), LAM (Marseille, France), LESIA (Paris, France), Laboratoire Lagrange (Nice, France), INAF–Osservatorio di Padova (Italy), Observatoire de Gen\`{e}ve (Switzerland), ETH Zurich (Switzerland), NOVA (Netherlands), ONERA (France) and ASTRON (Netherlands) in collaboration with ESO. SPHERE was funded by ESO, with additional contributions from CNRS (France), MPIA (Germany), INAF (Italy), FINES (Switzerland) and NOVA (Netherlands).  SPHERE also received funding from the European Commission Sixth and Seventh Framework Programmes as part of the Optical Infrared Coordination Network for Astronomy (OPTICON) under grant number RII3-Ct-2004-001566 for FP6 (2004-–2008), grant number 226604 for FP7 (2009-–2012) and grant number 312430 for FP7 (2013-–2016). We also acknowledge financial support from the Programme National de Plan\'{e}tologie (PNP) and the Programme National de Physique Stellaire (PNPS) of CNRS-INSU in France. This work has also been supported by a grant from the French Labex OSUG@2020 (Investissements d’avenir – ANR10 LABX56). The project is supported by CNRS, by the Agence Nationale de la Recherche (ANR-14-CE33-0018). It has also been carried out within the frame of the National Centre for Competence in Research PlanetS supported by the Swiss National Science Foundation (SNSF). MRM, HMS, and SD are pleased  to acknowledge this financial support of the SNSF. Finally, this work has made use of the the SPHERE Data Centre, jointly operated by OSUG/IPAG (Grenoble), PYTHEAS/LAM/CESAM (Marseille), OCA/Lagrange (Nice), Observatoire de Paris/LESIA (Paris), and Observatoire de Lyon, also supported by a grant from Labex  OSUG@2020 (Investissements d’avenir – ANR10 LABX56). We thank P. Delorme and E. Lagadec (SPHERE Data Centre) for their efficient help during the data reduction process.
   
The authors express their sincere gratitude to Tracy Beck for the constructive review, which greatly helped to improved the manuscript, and to the LBT AO and the LUCI AO commissioning teams for the support and observing T Tauri as one of their commissioning targets. The LBT is an international collaboration among institutions in the United States, Italy and Germany. The LBT Corporation partners are: The University of Arizona on behalf of the Arizona university system; Instituto Nazionale di Astrofisica, Italy; LBT Beteiligungsgesellschaft, Germany, representing the Max Planck Society, the Astrophysical Institute Potsdam, and Heidelberg University; The Ohio State University; The Research Corporation, on behalf of The University of Notre Dame, University of Minnesota and University of Virginia. IRAF is distributed by the National Optical Astronomy Observatory, which is operated by the Association of Universities for Research in Astronomy (AURA) under cooperative agreement with the National Science Foundation \citep{IRAFcite}.

FMe acknowledges funding from ANR of France under contract number ANR-16-CE31-0013.
\end{acknowledgements}

%
%

\bibliographystyle{aa}
\bibliography{kasperrefs}

\end{document}